\documentclass[prd,notitlepage,showpacs,nofootinbib,superscriptaddress]{revtex4-2}

\usepackage[utf8]{inputenc} 

\usepackage{amsmath}
\usepackage{amssymb}
\usepackage{float}
\usepackage{comment}
\usepackage{slashed}
\usepackage[normalem]{ulem}
\usepackage{bbold}
\usepackage{cancel}
\usepackage{graphicx}
\usepackage{multirow}
\usepackage{rotating}
\usepackage{caption}
\usepackage{subcaption}
\usepackage{tabulary}

\usepackage{array}
\newcolumntype{P}[1]{>{\centering\arraybackslash}p{#1}}
\def\gsim{\raise0.3ex\hbox{$\;>$\kern-0.75em\raise-1.1ex\hbox{$\sim\;$}}}
\def\lsim{\raise0.3ex\hbox{$\;<$\kern-0.75em\raise-1.1ex\hbox{$\sim\;$}}}

\usepackage[usenames,dvipsnames]{color}

\newcommand {\ignore}[1]{}

\usepackage[colorinlistoftodos]{todonotes}
\usepackage[colorlinks=true,linkcolor=darkblue,citecolor=darkblue,urlcolor=darkblue, pdfborder={0 0 0}]{hyperref}
\usepackage[normalem]{ulem}

\usepackage{caption}
\usepackage{subcaption}
\definecolor{linkcolor}{rgb}{0,0,0.8}

\definecolor{darkgreen}{rgb}{0,0.5,0}
\definecolor{darkred}{rgb}{0.6,0,0}
\definecolor{brown}{rgb}{0.59, 0.29, 0.0}
\definecolor{mightnightblue}{RGB}{25,25,112}
\definecolor{darkblue}{rgb}{0,0,0.8}

\newcommand {\darkblue} {\color{darkblue}}

\def\EW{$\mathrm{SU(2)_L \otimes U(1)_Y}$ }

\def\Neff{N_{\rm eff}}

\newcommand{\AddrCFTP}{%
Departamento de F\'{\i}sica and CFTP, Instituto Superior T\'ecnico, Universidade de Lisboa, Av. Rovisco Pais 1, 1049-001 Lisboa, Portugal}

\newcommand{\AddrNCBJ}{%
National Centre for Nuclear Research, Pasteura 7, Warsaw, PL-02-093, Poland}

\usepackage[force]{feynmp-auto}

\begin{document}

\title{\darkblue Sterile neutrino Dark Matter in the minimal Dirac Seesaw}

\author{J. Adhikary}\email{jyotismita.adhikary@ncbj.gov.pl}
\affiliation{\AddrNCBJ}
\author{A. Batra}\email{aditya.batra@tecnico.ulisboa.pt}
\affiliation{\AddrCFTP}
\author{K. Deka}\email{kishan.deka@ncbj.gov.pl}
\affiliation{\AddrNCBJ}
\author{ F.~R. Joaquim}\email{filipe.joaquim@tecnico.ulisboa.pt}
\affiliation{\AddrCFTP}

\begin{abstract}
\vspace{+0.1cm}
We study sterile neutrino dark matter in a minimal Type-I Dirac seesaw framework where the states responsible for generating Dirac neutrino masses at tree level can be viable dark matter candidates. A $\mathcal{Z}_6$ symmetry, spontaneously broken to a residual $\mathcal{Z}_3$ by the vacuum expectation value of a singlet scalar, forbids Majorana mass operators and ensures neutrino Diracness. The lightest sterile neutrino is produced non-thermally via freeze-in from decays of Standard Model particles and an additional scalar state. We show that the presence of an additional right-handed mixing angle, $\theta_R$, opens up viable regions of parameter space where the observed dark matter relic abundance can be reproduced while maintaining cosmological stability. This mainly stems from the absence of X-ray astrophysical constraints in our scenario. We further find that the freeze-in production of right-handed neutrinos yields a negligible contribution to $\Delta N_{\rm
eff}$, consistent with current cosmological bounds.
\end{abstract}

\maketitle
\noindent

\section{Introduction}
\label{sec:intro}
The Standard Model (SM) of particle physics provides a remarkably accurate description of fundamental interactions and has been validated by a wide range of experimental results. Nevertheless, it is widely recognized as an incomplete theory, as it does not incorporate mechanisms to explain two empirically established phenomena: the generation of non-zero neutrino masses~\cite{Kajita:2016cak,McDonald:2016ixn} and the existence of cosmologically stable dark matter (DM)~\cite{Bertone:2004pz,Planck:2018vyg}. A minimal and theoretically well-motivated extension to address the first issue is the Type-I seesaw mechanism~\cite{Minkowski:1977sc,Gell-Mann:1979vob,Yanagida:1979as,Schechter:1980gr,Glashow:1979nm,Mohapatra:1979ia}, which introduces heavy right-handed (sterile) neutrinos. Through their interaction with the SM, these states generate naturally small masses for the active neutrinos via the seesaw relation. Within this same framework, the lightest sterile neutrino can also account for the observed DM relic abundance. In this case, its production occurs via the freeze-in mechanism, characterizing it as a Feebly Interacting Massive Particle (FIMP) dark matter candidate.

In the minimal scenario, sterile neutrinos are produced non-thermally in the early Universe via their mixing with SM active neutrinos. This process, known as the Dodelson–Widrow (DW) scenario~\cite{Dodelson:1993je}, arises from the small but finite probability for active neutrinos in the thermal plasma to oscillate into sterile states. As the Universe expands and cools, this out-of-equilibrium production gradually builds up a relic population of sterile neutrinos. The resulting cosmological abundance is controlled primarily by two parameters: the sterile neutrino mass and the active–sterile mixing angle. Reproducing the observed dark matter relic density inferred by the Planck satellite, $0.1126 < \Omega h^2 < 0.1246$~\cite{Planck:2018vyg}, typically requires sterile neutrino masses in the keV range and highly suppressed mixing angles.

However, this minimal framework is subject to severe phenomenological constraints. The same mixing that enables sterile neutrino production also induces radiative decays of the form $N \to \nu \gamma$~\cite{Pal:1981rm}. Although the corresponding lifetime typically exceeds the age of the Universe, these decays would give rise to a monochromatic X-ray line potentially observable in astrophysical data. The absence of such signals in current X-ray observations places stringent upper bounds on the active–sterile mixing angle~\cite{Boyarsky:2007ge,Horiuchi:2013noa,Roach:2019ctw,Foster:2021ngm,Malyshev:2020hcc,Dekker:2021bos,Ando:2021fhj}. When combined with independent constraints from structure formation~\cite{Tremaine:1979we,Boyarsky:2008ju,Merle:2015vzu,Abazajian:2017tcc,Yeche:2017upn}, these limits effectively exclude the region of parameter space in which the DW scenario accounts for the entire dark matter abundance.

Despite these challenges, sterile neutrinos remain attractive dark matter candidates due to the simplicity of the underlying framework. This has motivated the exploration of alternative production scenarios. A well-known example is the Shi–Fuller mechanism~\cite{Shi:1998km}, where resonant active–sterile oscillations occur in the presence of a large lepton asymmetry in the early Universe. This mechanism relaxes the constraints on the mixing angle, but typically requires a significant degree of degeneracy among heavy sterile neutrinos. Other possibilities include scenarios with neutrino self-interactions~\cite{DeGouvea:2019wpf,Berryman:2022hds,Astros:2023xhe}, production from the decay of singlet scalar fields~\cite{Petraki:2007gq,Merle:2013wta,Adulpravitchai:2014xna,Merle:2015oja}, and non-standard cosmological histories with late-time entropy injection that modify the relic abundance~\cite{Hansen:2017rxr}.

Within the conventional Type-I seesaw, sterile neutrinos can also be produced through the decays of SM gauge bosons and the Higgs boson~\cite{Datta:2021elq}. The corresponding production rates are governed by Yukawa couplings, which simultaneously set the active–sterile mixing angles responsible for their radiative decays. Requiring the DM candidate to remain cosmologically stable, together with the stringent bounds from X-ray observations, typically forces these Yukawa couplings to be extremely small, strongly suppressing this production channel. Nevertheless, thermal effects in the plasma can partially enhance the production rates~\cite{Abada:2023mib}.

An alternative possibility is that neutrinos are Dirac rather than Majorana particles. The continued non-observation of neutrinoless double beta decay ($0\nu\beta\beta$)~\cite{Jones:2021cga,Cirigliano:2022oqy,Dolinski:2019nrj} may point in this direction, as suggested by the black-box theorem~\cite{Schechter:1981bd}. In this case, neutrino mass generation requires mechanisms that forbid Majorana mass terms, typically through additional symmetries~\cite{Aranda:2013gga}. Such constructions include Dirac realizations of the Type-I~\cite{Ma:2014qra,Addazi:2016xuh,CentellesChulia:2016rms,Bonilla:2017ekt} and Type-II~\cite{Valle:2016kyz,Reig:2016ewy,Bonilla:2016zef} seesaw mechanisms. Moreover, due to the presence of additional parameters, Dirac seesaws do not typically require extremely large mediator masses, unlike Majorana seesaw models, to ensure the smallness of neutrino masses. Hence, having sterile neutrino masses at the scale required to explain DM ($\sim 100$ keV) is more natural. A particularly interesting feature of Dirac seesaw models is the presence of an additional mixing angle, $\theta_R$, associated with the right-handed neutrino sector. This parameter can contribute to sterile neutrino production via Higgs decays without directly affecting the radiative decay channel probed by X-ray observations. 

In this work, we propose a minimal version of the Type-I Dirac seesaw and demonstrate that the additional degree of freedom $\theta_R$ significantly relaxes X-ray constraints and opens up viable regions of parameter space for sterile neutrino dark matter. We further show that this setup provides a minimal realization of Dirac neutrino mass generation ``seeded'' by dark matter: the mechanism operates already at tree level, and is therefore structurally simpler than radiative constructions such as the Dirac scotogenic model~\cite{Bonilla:2018ynb}. This paper is organised as follows. In Sec.~\ref{sec:model} we introduce a minimal realization of the Dirac seesaw model. The phenomenological implications for dark matter and cosmology are discussed in Sec.~\ref{sec:pheno}. Finally, our conclusions are presented in Sec.~\ref{sec:concl}. Additional details on the scalar sector, and relevant formulas used in the calculations are provided in the appendices.

\section{The Model}
\label{sec:model}
We extend the SM by introducing three right-handed neutrinos $\nu_{Ri}$, three pairs of Dirac fermion fields $N_i$ and a real singlet $\sigma$. In addition, a $\mathcal{Z}_6$ symmetry is imposed to protect the Dirac nature of neutrinos. Assigning the $\mathcal{Z}_6$ charges as shown in Table~\ref{tab:model}, we forbid all effective Majorana neutrino mass generation operators of the type $(\bar{\ell_L^c} \tilde{\Phi}^\ast) (\tilde{\Phi}^\dagger \ell_L) \sigma^n \sigma^{\ast n^\prime} $, where $\ell_L=(\nu_L\;,\;e^-_L)^T$ and $\Phi=(\phi^+\;,\;\phi^0)^T$ denote the SM lepton and Higgs doublet, respectively, and $\tilde{\Phi}=i \tau_2 \Phi$, with $\tau_2$ being the complex Pauli matrix. $n=n^\prime=0$ corresponds to the well-known Weinberg operator~\cite{Weinberg:1980wa}. After $\sigma$ gains a vacuum expectation value (VEV)~\footnote{The scalar sector is discussed in detail in Appendix~\ref{sec:scalar}, where we present the scalar mass spectrum and mixing structure, together with the vacuum stability conditions imposed in our numerical analysis.}, the $\mathcal{Z}_6$ symmetry is spontaneously broken to a residual unbroken $\mathcal{Z}_3$ symmetry. The fields that have non-trivial charges under the $\mathcal{Z}_3$ symmetry transform as: $(\ell_L,e_R,\nu_R,N_{L,R}) \to \omega (\ell_L,e_R,\nu_R,N_{L,R})$. This $\mathcal{Z}_3$ symmetry forbids all the above Majorana type operators. The idea of using residual $\mathcal{Z}_N$ symmetries to prevent Majorana neutrino masses has previously been explored in the context of a $\text{U}(1)_{B-L}$ symmetry in Ref.~\cite{Bonilla:2018ynb}. Effective Dirac neutrino mass operators, generically expressed as, $ (\bar{\ell_L} \tilde{\Phi} \nu_R)\, \sigma^n \sigma^{\ast n^\prime} $ are allowed by the $\mathcal{Z}_3$ symmetry, with the lowest dimension operator being the dimension-5 $(\bar{\ell_L} \tilde{\Phi} \nu_R)\sigma$, corresponding to $n=1,n^\prime=0$. The tree-level Yukawa interaction $\bar{\ell_L} \tilde{\Phi} \nu_R$ is forbidden by the $\mathcal{Z}_6$ symmetry.

\begin{table}[t!]
	\centering
	\begin{tabular}{| c | c | c | c | }
		\hline 
& \; Fields \; & \; \EW \; &  \; $\mathcal{Z}_6$ \;  \\
		\hline 
		\multirow{4}{*}{\; Fermions \;} 
&$\ell_L$&($\mathbf{2}, {-1/2}$)& $\omega$   \\
&$e_R$&($\mathbf{1}, {-1}$)& {$\omega$}   \\
&$\nu_R$&($\mathbf{1}, {0}$)& {$\omega^4$}    \\
&$N_{L,R}$&($\mathbf{1}, {0}$)& {$\omega$}   \\
		\hline 
\multirow{2}{*}{Scalars}  &$\Phi$&($\mathbf{2}, {1/2}$)& {$1$}    \\
&$\sigma$&($\mathbf{1},0$)& {$\omega^3$}    \\	
\hline
	\end{tabular}
	\caption{Field content and transformation properties under \EW and $\mathcal{Z}_6$.}
	\label{tab:model} 
\end{table}

With the field content and symmetries defined above, the Yukawa Lagrangian takes the form,
\begin{align}
    -\mathcal{L} = \mathbf{Y}_e \overline{\ell_L} \Phi e_R + \mathbf{Y}_\nu \overline{\ell_L} \tilde{\Phi} N_R + \mathbf{Y}_\sigma \overline{N_L} \nu_R \sigma + \mathbf{M}_N \overline{N_L} N_R + \text{H.c.} \; ,
    \label{eq:LYuk}
\end{align}
where the Yukawa couplings $\mathbf{Y}_e$, $\mathbf{Y}_\nu$ and $\mathbf{Y}_\sigma$ and the bare mass $\mathbf{M}_N$ are $3 \times 3$ matrices. 
\begin{figure}[t!]
    \centering
    \includegraphics[scale=0.9]{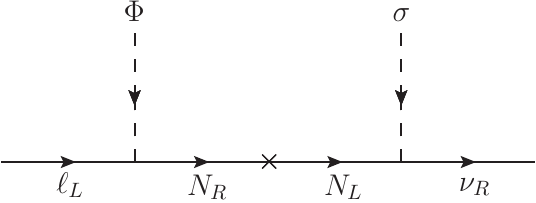} 
    \caption{Feynman diagram for tree-level Dirac neutrino mass generation with U$(1)_L$ spontaneously broken.}
\label{fig:neutrino}
\end{figure}
These couplings induce nonzero neutrino masses through a Dirac seesaw, as illustrated in Fig.~\ref{fig:neutrino}.
The $6\times6$ Dirac neutrino mass matrix can be written in the basis $(\nu_L,N_L,\nu_R,N_R)^T$ as,
\begin{align}
\mathcal{M}_{\nu,N} = \begin{pmatrix}
 0 & \mathbf{Y}_\nu v_\phi/\sqrt{2}  \\
 \mathbf{Y}_\sigma v_\sigma/\sqrt{2} & \mathbf{M}_N  \\
\end{pmatrix} \; ,
\end{align}
where $v_\phi$ and $v_\sigma$ are the VEVs of the Higgs field $\Phi$ and the real singlet $\sigma$ respectively. In the limit $M_{N_{ij}} \gg v_\phi,v_\sigma$, the effective $3\times3$ mass matrix for the active neutrinos is approximately given by~\cite{Schechter:1981cv},
\begin{align}
\mathbf{M}_\nu \simeq \frac{v_\phi v_\sigma}{2} \mathbf{Y}_\nu \mathbf{M}^{-1} \mathbf{Y}_\sigma \;.
\end{align}
The rotation to the neutrino mass-eigenstate basis can be defined as,
\begin{equation}
(\nu_e,\nu_\mu,\nu_\tau,N_e,N_\mu,N_\tau)^T_{L,R}
 = \mathbf{U}_{L,R}
(\nu_1,\nu_2,\nu_3,N_1,N_2,N_3)^T_{L,R}\; ,
\label{eq:neumixing}
\end{equation}
with $\nu_i$ ($N_i$) denoting the physical active (sterile) Dirac neutrino states. For the sake of simplicity, we neglect intergenerational mixing among active and sterile neutrinos. In this case, the left and right active–sterile mixing can be parametrised by the angles $\theta_{Li}$ and $\theta_{Ri}$, respectively. For each generation, these are given as,
\begin{align}
\tan{(2\theta_{Li})}=\frac{2\sqrt{2} M_{N_i} Y_{\nu_i} v_\phi}{Y_{\sigma_i} v_\sigma^2 + 2 M_{N_i}^2 - Y_{\nu_i} v_\phi^2} \; \;,
\;\;
\tan{(2\theta_{Ri})}=\frac{2\sqrt{2} M_{N_i} Y_{\sigma_i} v_\sigma}{Y_{\nu_i} v_\phi^2 + 2 M_{N_i}^2 - Y_{\sigma_i} v_\sigma^2} \;\;  \;\;(i=1,2,3)\,,
\label{eq:tan2theta}
\end{align}
where the $Y$'s and $M_N$'s are the diagonal elements of the corresponding matrices. A more general treatment of the Yukawa and mass matrices would provide sufficient freedom to reproduce the observed neutrino mass splittings and leptonic mixing angles measured in oscillation experiments while leading to the same conclusions regarding DM phenomenology (see Sec.~\ref{sec:DM}). A complete study of flavour textures and oscillation-fit predictions would require a dedicated analysis and is beyond the scope of this work.

Before moving on to the phenomenology of the model, it is worth commenting on the choice of the number of species of $N_{L,R}$ and $\nu_R$.  In the Dirac seesaw, it is possible to correctly reproduce the observed neutrino oscillation data with only two species of $N_{L,R}$ and $\nu_R$ each. This would lead to one massless active neutrino. However, the experimentally measured solar mass-squared difference $\Delta m^2_{12}$ would place a very tight constraint on the mass of the lightest sterile neutrino and on the mixing angles $\theta_{L1,R1}$. In fact, the resulting available parameter space would already be excluded by X-ray searches of the radiative decay of a sterile neutrino $N \to \nu \gamma$ which highly constrains the mass and mixing angle of the sterile neutrino (see Sec.~\ref{sec:DM} for details). Therefore, in this work, we consider three species of $N_{L,R}$ and $\nu_R$ each. 
The masses and Yukawa couplings associated with the second and third generations of sterile neutrinos are not constrained by the dark matter requirements and can be chosen to fit the oscillation data.

\section{Phenomenology}
\label{sec:pheno}
In this section, we study the phenomenological implications of our model for DM and the effective number of relativistic degrees of freedom, $N_\text{eff}$. We perform a numerical scan over the mixing angles $\theta_{L1,R1}$ and the mass of the dark matter candidate $M_{N_1}$. The SM Higgs boson mass is fixed to its experimentally measured value $m_h=125.20\,\mathrm{GeV}$~\cite{ParticleDataGroup:2024cfk}, and we choose $m_H=500\,{\rm GeV}$, $\alpha=0.1$, and $v_\sigma=150\,{\rm GeV}$. The scalar couplings are determined by eqs.~\eqref{eq:lambda} given in Appendix~\ref{sec:scalar}, where we also present the scalar mass spectrum and mixing, together with the vacuum stability conditions imposed in our numerical analysis. For simplicity, we assume the Yukawa couplings to be real and extract their values using eq.~\eqref{eq:tan2theta}. We have chosen these benchmark parameters to clearly illustrate the interplay between the left- and right-handed mixing angles in the DM dynamics (see Sec.~\ref{sec:DM}), while ensuring that the scalar potential remains bounded from below, all scalar couplings are perturbative, and the mixing between the new scalar state and the Higgs boson is sufficiently small to evade current collider constraints ($\sin{\alpha} \lesssim 0.1$). As a result of this mixing, the new scalar $H$ inherits suppressed couplings to SM particles proportional to $\sin\alpha$. Consequently, it can be produced at collider experiments through the same mechanisms as the SM Higgs boson, albeit with reduced production cross sections. Depending on its mass, the scalar may be probed through direct searches in standard Higgs decay channels as well as through precision measurements of Higgs signal strengths~\cite{Fernandez-Martinez:2022stj}.

\subsection{Dark matter relic abundance}
\label{sec:DM}
In the present framework, the lightest sterile neutrino $N_1$ is very weakly coupled to the SM and is a viable DM candidate through a freeze-in mechanism. Its production occurs via the following dominant decay processes:
\begin{align}
    W^+ \to N_1 e^+_i, \;\;\; 
    W^- \to \overline{N}_1 e^-_i, \;\;\; 
    Z \to N_1 \overline{N}_1, \;\;\; 
    h/H \to N_1 \overline{\nu}_1, \;\;\; 
    h/H \to \overline{N}_1 \nu_1, \;\;\; 
\label{eq:process}
\end{align}
where $h$ and $H$ are the physical scalars explicitly defined in Appendix~\ref{sec:scalar}. In our Dirac seesaw model, the interactions involving these scalars stem from the Yukawa terms $\mathbf{Y}_\sigma \overline{N}_L \nu_R \sigma$ and $\mathbf{Y}_\nu \overline{\ell_L} \tilde{\Phi} N_R$, which, in the mass basis read: 
\begin{align}
    -\mathcal{L}_\text{int} = \frac{1}{\sqrt{2}}\sum_{j,l=1}^{6}\sum_{i=1}^{3}\sum_{k=4}^{6} \left[
    \mathbf{U}^{\ast ji}_L \mathbf{U}^{kl}_R (h c_{\alpha} + H s_{\alpha}) \mathbf{Y}_\nu^{i,k-3} + \mathbf{U}^{\ast jk}_L \mathbf{U}^{il}_R (-h s_{\alpha} + H c_{\alpha}) \mathbf{Y}_\sigma^{k-3,i}\right] \overline{\nu}_{jL} \nu_{lR}
    + \text{H.c.} \; ,
    \label{eq:Lint}
\end{align}
with $\mathbf{U}_{L,R}$ as defined in eq.~\eqref{eq:neumixing} and $\alpha$ being the scalar mixing angle given in eq.~\eqref{eq:scalarmixing}. From now on, $\nu_{1-3}$ ($\nu_{4-6}$) denote the active (sterile) Dirac neutrino mass eigenstates. Thus, fixing $j,l=4$ leads to terms involved in the production of  the lightest sterile neutrino $N_1$. 

Besides the freeze-in decays given in eq.~\eqref{eq:process}, there can also be contribution to the production of $N_1$ from the DW mechanism, namely \cite{Abazajian:2001nj},
\begin{equation}
\Omega_{\text{DW}} h^{2} \simeq 0.3 \times 
\left[ \frac{\sin^{2} (2\theta_{L1})}{10^{-10}} \right]
\left( \frac{m_{s}}{100 \, \text{keV}} \right)^{2},
\end{equation}
where $m_s$ corresponds here to the mass of $N_1$, i.e. $m_s = M_{\nu_4} \simeq M_{N_1}$. In our case, to obtain the correct observed relic density, $\theta_{L1}$ must be $\sim 10^{-9}$ when $m_s \sim 10^{-4} \text{ GeV}= 100$ keV. For these parameter choices, the corresponding DW contribution to the relic density is $\sim 10^{-8}$ being, therefore, negligible. 

The way we compute the relic abundance of the sterile fermion $N_1$ is by numerically solving the Boltzmann equation for its comoving number density,
\begin{align}
\frac{dY_{N_1}}{dz}
&=
\frac{2 M_{\rm Pl}}{1.66\, m_h^2}\,
\frac{z\, \sqrt{g_*(T)}}{g_{*s}(T)}
\sum_i
\left\langle \Gamma_i \right\rangle
\left(
Y_i^{\rm eq} - Y_{N_1}
\right),
\label{eq:Boltzmann}
\end{align}
where $z \equiv m_h/T$, $g_*(T)$ and $g_{*s}(T)$ denote the effective relativistic degrees of freedom for the energy and entropy densities, respectively. The expressions for the thermally-averaged decay widths $\left\langle \Gamma_i \right\rangle$ for the various modes were taken from \cite{Biswas:2016bfo}. In the case of $h$ and $H$ decays, the final states are $N_1 \overline{\nu}_1$ and $\overline{N}_1 \nu_1$ instead of $N_{1}\overline{N}_{1}$.  
Relevant $2\to 2$ processes, such as $hh \to N_1\overline{\nu}_1$ and $HH \to N_1\overline{\nu}_1$ proceeding via $s$-channel scalar exchange, involve two heavy initial-state particles whose thermal number densities are Boltzmann-suppressed below their respective mass thresholds. Since the $2\to 2$ rate is proportional to the product of two such densities, the suppression is squared relative to that of the $1\to 2$ decay channels. The annihilation amplitude involves the product of a trilinear scalar coupling and a Yukawa coupling, whereas the decay amplitude involves only the Yukawa coupling; since the relevant Yukawa couplings are of order $\theta_{L,R}$ in the freeze-in regime, this additional factor further suppresses the annihilation contribution. Furthermore, there is an additional suppression of the fourth power of the mediator mass ($ \sim m^4$) compared to decay channels. We have verified numerically that the contribution of these channels to $\Omega_{N_1} h^2$ is of order $\mathcal{O}(10^{-24})$, confirming their irrelevance across the full parameter space of interest. The equilibrium yield for a generic particle species $i$ of mass $m_i$ and internal degrees of freedom $g_i$ is taken to be,
\begin{equation}
Y_i^{\rm eq}(z)
=
\frac{45}{4\pi^4}
\frac{g_i}{g_{*s}(T)}
\left( \frac{m_i}{m_h} z \right)^2
K_2\!\left( \frac{m_i}{m_h} z \right).
\label{eq:Yeq_generic}
\end{equation}
The Boltzmann equation~\eqref{eq:Boltzmann} is solved numerically and the final yield at late times is converted into the present-day relic density via
\begin{equation}
\Omega_{N_1} h^2 = 2.755 \times 10^8 \, \left(\frac{M_{N_1}}{\mathrm{GeV}}\right) Y_{N_1}(T_{present})     \ ,
\end{equation}
where, $Y_{N_1}(T_{present})$ represents the comoving density of DM at present time. In our numerical scan, we apply the following constraints: 
\begin{itemize}
    \item Perturbativity of scalar couplings ($\lambda_\Phi, \lambda_\sigma, \lambda_{\Phi\sigma} \le 4\pi$) and Yukawa couplings ($Y_{\nu_1}, Y_{\sigma_1} \le \sqrt{4\pi}$).
    \item Sum of neutrino masses $\le 0.12\,{\rm eV}$, a constraint obtained from the observations of the CMB, combined with lensing and baryon acoustic oscillations 
    data~\cite{ParticleDataGroup:2024cfk}.
    \item Non-thermality condition: the rate of production of $N_1$ is less than the expansion rate of the Universe at a temperature $T \sim M$, where $M$ is the mass of the decaying particle~\cite{Arcadi:2013aba}. That is, $\Gamma/H < 1$, where $\Gamma$ is the decay width and $H$ is the Hubble parameter. 
    \item The sterile neutrino $N_1$ can decay into SM fermions depending on its mass. Namely:
    \begin{itemize}
        \item For $m < 2 m_e$, the dominant decay channel is $N_1 \to \nu_1 \nu_i \overline{\nu}_i$, where $i=1,2,3$. This decay is mediated by virtual $Z$, $h$ and $H$ bosons. The $h$ and $H$ mediated decay widths are much suppressed compared to the $Z$~\cite{Boyarsky:2018tvu}.
        \item For $m \ge 2 m_e$, the decay $N_1 \to \nu_1 e^+ e^-$ opens up. This decay is mediated by virtual $W$, $Z$, $h$ and $H$ bosons. However, this decay width is much smaller compared to $N_1 \to \nu_1 \nu_i \overline{\nu}_i$.
        \item For $m \ge 100$ MeV, several possibilities open up, including hadronic decay modes~\cite{Atre:2009rg}, therefore we have considered sterile neutrino masses upto 100 MeV.
    \end{itemize}    
    We require these decays to be suppressed so that the lifetime of $N_1$ and $\overline{N}_1$ is longer than the age of the Universe, $t_\text{Universe} = 4.4 \times 10^{17}$ s. 
\end{itemize}

\begin{table}[t!]
\centering
\renewcommand{\arraystretch}{1.3}

\begin{tabular}{|c|c|c|}
\hline
\textbf{Decay channel} & \textbf{$\theta_L$-dominated} & \textbf{$\theta_R$-dominated} \\
\hline
$W^{\pm}$ & $ \sim 1$ & $\sim 0$ \\
\hline
$Z$ & $\sim 0$ & $\sim 0$ \\
\hline
$H$ & $ \sim 0$ & $ \sim 0.44$ \\
\hline
$h$ & $\sim 0$ & $ \sim 0.56$ \\
\hline
\end{tabular}
\caption{Relative contributions of individual decay channels to the DM relic density for two benchmark points, chosen to exhibit the maximum depletion of the relic abundance due to back-reaction effects, see Fig.~\ref{fig:freeze_in_evol} for the corresponding abundance evolution. The ``$\theta_L$-dominated'' column corresponds to $M_{N_1} = 9.98\times 10^{-4}\,\mathrm{GeV}$, $\theta_{L1} = 3.95 \times 10^{-9}$, and $\theta_{R1} = 10^{-10}$, yielding $\Omega h^2 = 0.1211$. The ``$\theta_R$-dominated'' column corresponds to $M_{N_1} = 9.98\times 10^{-2}\,\mathrm{GeV}$, $\theta_{R1} = 1.74 \times 10^{-6}$, and $\theta_{L1} = 10^{-15}$, yielding $\Omega h^2 = 0.1128$. Here, $W^\pm$ decays are entirely absent since $\theta_L$ is negligible, and DM is produced exclusively through the scalar sector, with $h$ and $H$ decays contributing approximately $56\%$ and $44\%$ respectively.
}
\label{tab:relic_channels}
\end{table}

\begin{figure}[ht!]
\centering
\includegraphics[width=0.49\linewidth]{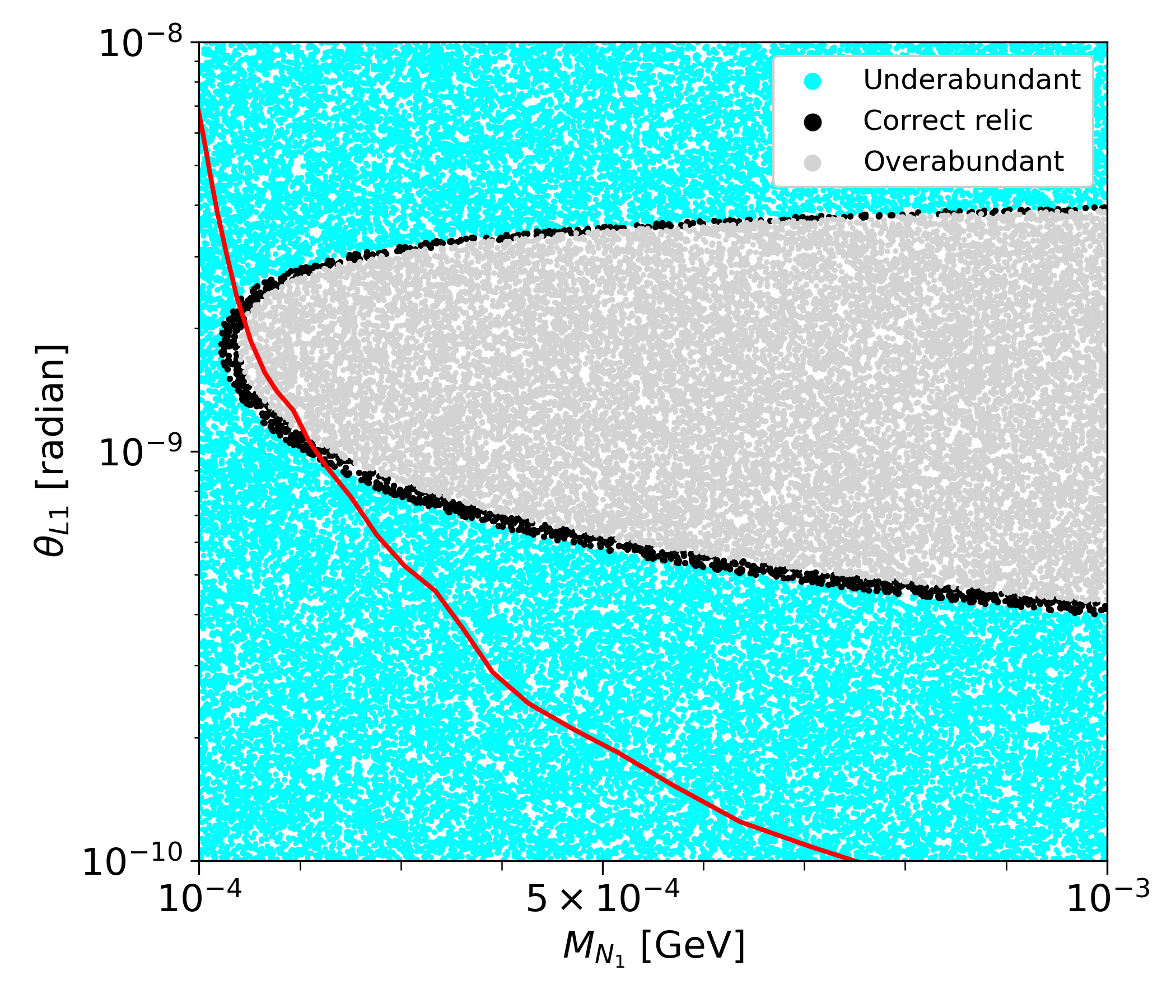}
\includegraphics[width=0.49\linewidth]{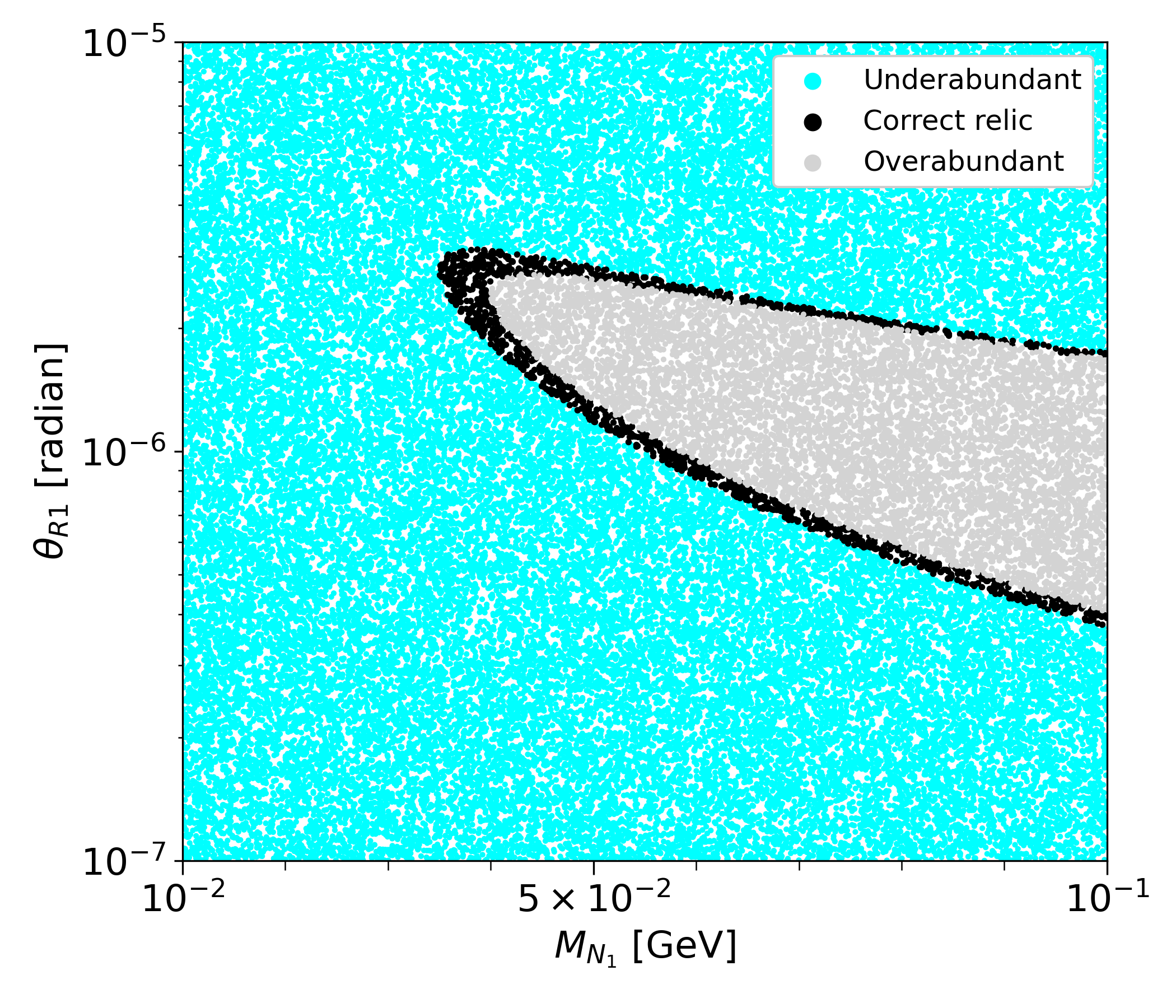}
\caption{Viable parameter space for sterile neutrino dark matter from our numerical scan. Blue, black, and grey points correspond to underabundant ($\Omega h^2 < 0.1126$), cosmologically viable ($0.1126 \leq \Omega h^2 \leq 0.1246$), and overabundant ($\Omega h^2 > 0.1246$) regions, respectively, at the $3\sigma$ level inferred 
from Planck~\cite{Planck:2018vyg}. \textit{Left:} Parameter space in the $(M_{N_1}\,[\mathrm{GeV}],\,\theta_{L1})$ plane with $\theta_{R1} = 10^{-10}$ fixed, so that production is entirely controlled by the left-handed mixing angle. The solid red curve indicates the upper bound on $\theta_{L1}$ from the non-observation of the radiative decay $N_1 \to \nu \gamma$ in X-ray astrophysical surveys~\cite{Boyarsky:2007ge,Horiuchi:2013noa,Roach:2019ctw,Foster:2021ngm}, with the region above it excluded. \textit{Right:} Parameter space in the $(M_{N_1}\,[\mathrm{GeV}],\,\theta_{R1})$ plane with $\theta_{L1} = 10^{-15}$ fixed, isolating the effect of the right-handed mixing angle. Since $\theta_{R1}$ does not enter the $W$-loop mediating $N_1 \to \nu \gamma$, the X-ray bound is entirely absent in this regime. As a consequence, viable DM masses extend to $M_{N_1} \gtrsim 10^{-2}\,\mathrm{GeV}$, demonstrating that the right-handed mixing angle opens up a qualitatively new and otherwise inaccessible region of parameter space.
}
\label{fig:Corr_MN1_theta}
\end{figure}
\begin{figure}[ht!]
\centering
\includegraphics[width=0.495\linewidth]{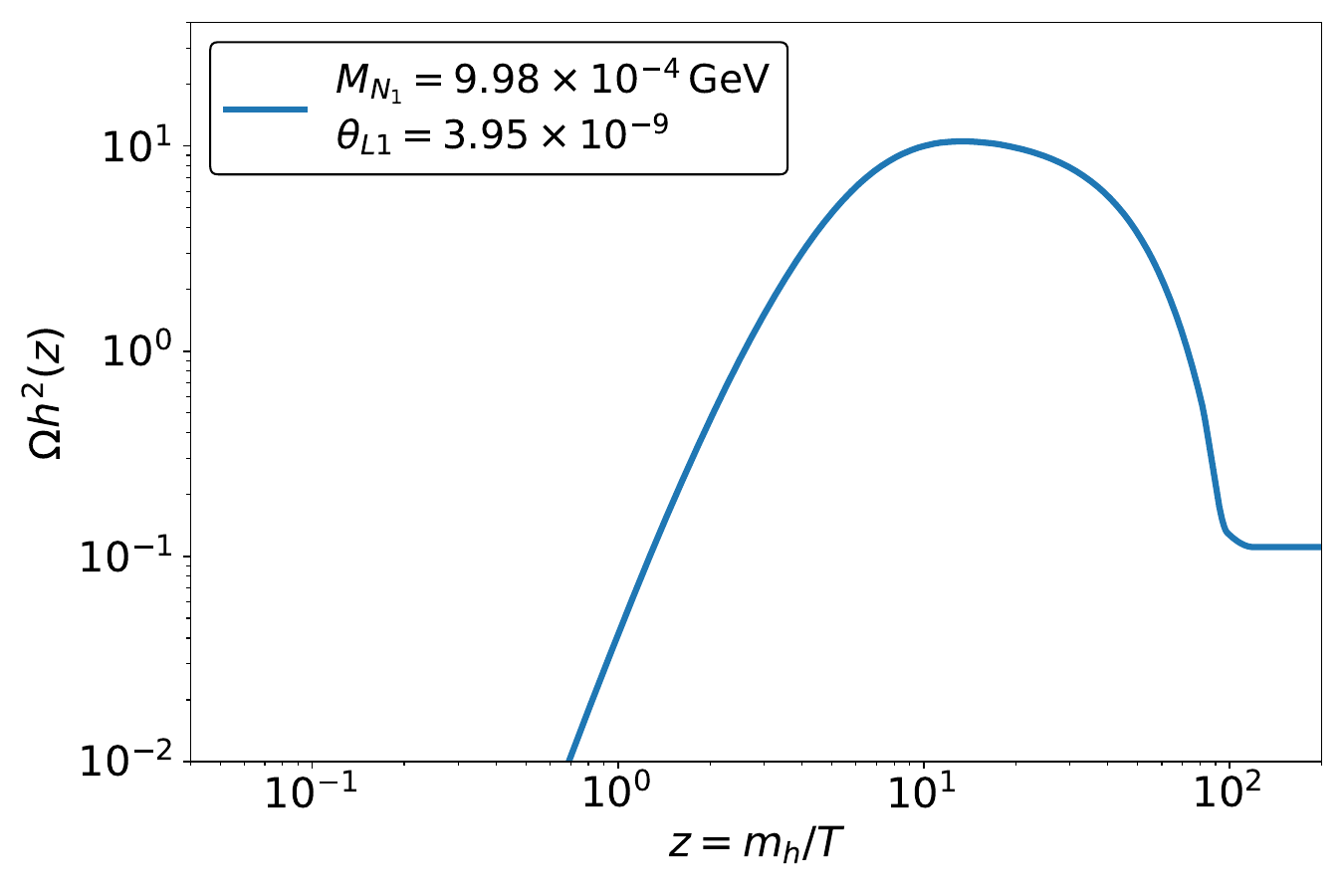}
\includegraphics[width=0.495\linewidth]{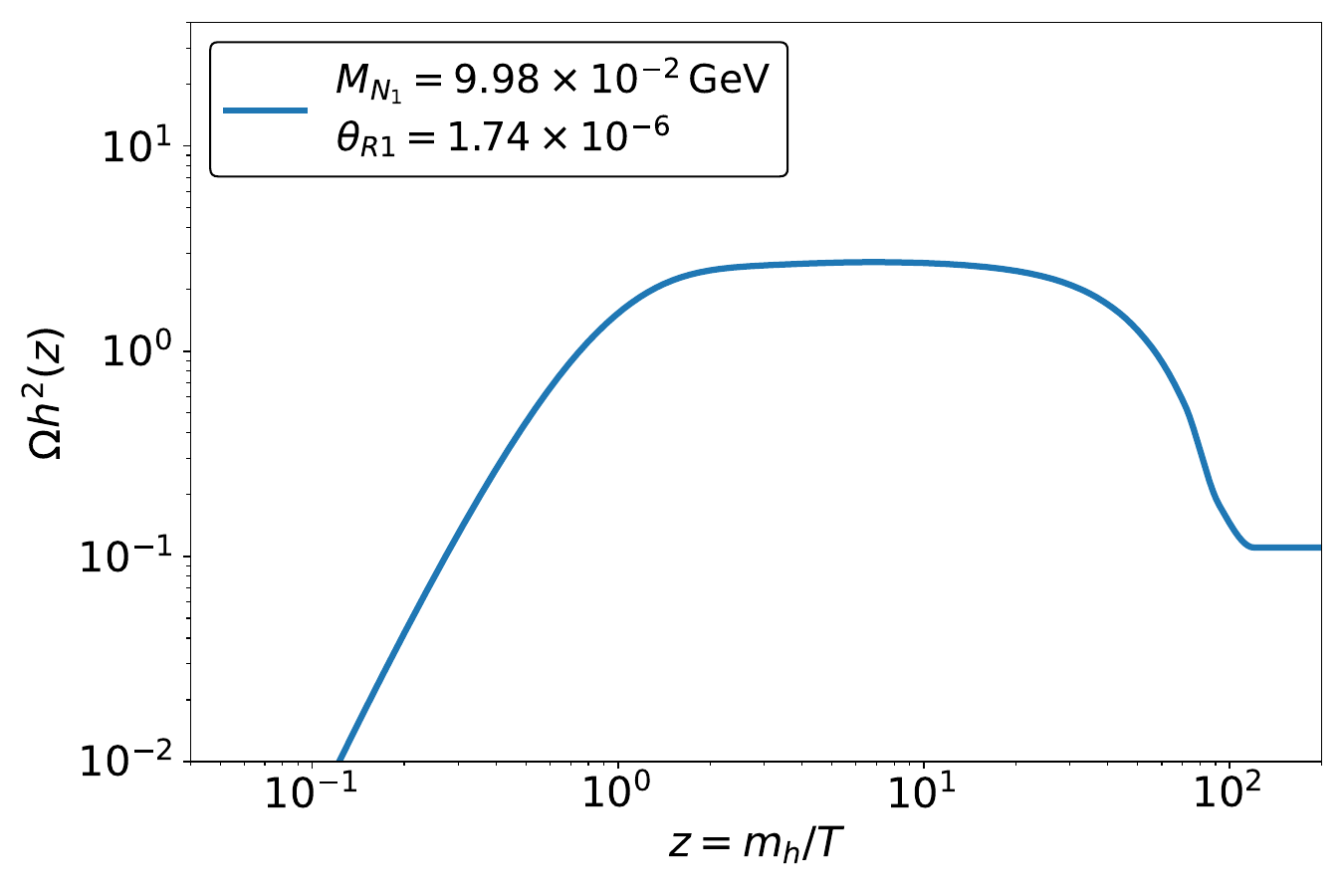}
\caption{Evolution of the DM relic abundance $\Omega h^2(z)$ as a function of 
$z \equiv m_h/T$ for two benchmark points selected from 
Fig.~\ref{fig:Corr_MN1_theta}, chosen to exhibit the largest fractional depletion of the peak abundance due to inverse processes within each production regime (see Table~\ref{tab:relic_channels} for the corresponding channel decomposition). \textit{Left:} $\theta_L$-dominated production, with $M_{N_1} = 9.98\times 10^{-4}\,\mathrm{GeV}$, $\theta_{L1} = 3.95 \times 10^{-9}$, and $\theta_{R1} = 10^{-10}$. \textit{Right:} $\theta_R$-dominated production, with $M_{N_1} = 9.98\times 10^{-2}\,\mathrm{GeV}$, 
$\theta_{R1} = 1.74 \times 10^{-6}$, and $\theta_{L1} = 10^{-15}$. 
}
\label{fig:freeze_in_evol}
\end{figure}
In Fig.~\ref{fig:Corr_MN1_theta} we show the results of our parameter scan. The blue points correspond to underabundant DM ($\Omega h^2 < 0.1126$), the grey points to overabundant ($\Omega h^2 > 0.1246$) and the black points are cosmologically viable ($0.1126 < \Omega h^2 < 0.1246$), as inferred from Planck satellite data at the $3 \sigma$ level~\cite{Planck:2018vyg}. In the left panel, we show the parameter space in the $(M_{N_1},\theta_{L1})$ plane fixing $\theta_R$ to a very small value, $\theta_R=10^{-10}$, so that DM is entirely produced through interactions involving $\theta_L$. The region above the red line is excluded by X-ray astrophysical searches. From Table~\ref{tab:relic_channels}, where the relative contribution from different decay  channels producing $N_1$ is shown for comparison, we can see that the dominant production channel is the $W^\pm$ decay. The results show that X-rays exclude almost the entire DM-allowed parameter space with masses larger than $\sim 10^{-4}$ GeV.
The right panel shows the parameter space in the $(M_{N_1},\theta_{R1})$ plane choosing $\theta_L=10^{-15}$. In this case, DM is entirely produced through interactions involving $\theta_R$ (see Table~\ref{tab:relic_channels}), $h$ and $H$ decays being the dominant production channels. These are usually suppressed in Majorana sterile neutrino DM models due to the tight constraint on the mixing angle. However, in our Dirac seesaw framework, these decays can be enhanced by $\theta_R$. Since the decay $N \to \nu \gamma$ occurs through a $W$ loop, $\theta_R$ remains unconstrained from X-ray observations. 
Sub-leading effects involving $\theta_R$ can arise through chirality-flipping mass insertions, induced left-handed mixings from mass matrix diagonalization, or higher-loop corrections. These contributions are suppressed by factors of $m_\nu/m_N\sim 10^{-9}-10^{-8}$ at the amplitude level, leading to decay-rate suppressions of order $10^{-18}-10^{-16}$, while two-loop and RG-induced effects are further suppressed to roughly $10^{-23}-10^{-21}$. Therefore, the dependence of the X-ray decay on the right-handed mixing angle is phenomenologically negligible. Hence, it is possible to obtain the correct DM relic density even for comparatively large mass scales. 
At these masses, sterile-neutrino dark matter evades the stringent constraints from cosmological observations of structure formation. In particular, phase-space density considerations in dwarf galaxies~\cite{Tremaine:1979we,Boyarsky:2008ju,Merle:2015vzu,Abazajian:2017tcc} impose a conservative lower bound of $m_N \gtrsim 2$ keV. This bound can be further strengthened by Lyman-$\alpha$ forest observations, potentially reaching $m_N\sim 30$ keV~\cite{Yeche:2017upn}.

In addition to the production channels, the Boltzmann equation also includes the corresponding inverse processes, which give rise to back-reaction effects. These appear through terms proportional to $(Y_i^{\mathrm{eq}} - Y_{N_1})$ for the decay processes. At early times, when DM abundance satisfies $Y_{N_1} \ll Y^{\mathrm{eq}}$, these contributions are positive and DM production proceeds efficiently through decays of particles in the thermal bath. As the abundance of $N_1$ increases, the corresponding inverse reactions, such as $N_1 + \nu \rightarrow h, H$ and $N_1 + \overline{\nu} \rightarrow h, H$, become progressively more relevant. These processes partially deplete the dark matter abundance and reduce the net production rate. As a consequence, the evolution of the relic abundance may exhibit a peak followed by a mild decrease before eventually approaching a constant value at late times. The peak typically appears at $z \sim \mathcal{O}(10$--$100)$, when the temperature of the thermal bath drops below the masses of the particles responsible for the dominant production channels. In this regime, the equilibrium number densities of these particles become Boltzmann suppressed, leading to a rapid reduction in the production rate. At the same time, inverse processes start to partially counteract the accumulated abundance, producing a small depletion before the yield ultimately freezes to its asymptotic value.

Two examples of the aforementioned behaviour are shown in Fig.~\ref{fig:freeze_in_evol}. In the left panel, the relic abundance is shown as a function of $z$, for a benchmark point from the left panel of Fig.~\ref{fig:Corr_MN1_theta} with a sizable $\theta_L$ and keeping $\theta_R = 10^{-10}$, illustrating the effect of the left-handed mixing on freeze-in production. Conversely, the right panel shows the evolution for a benchmark point from the right panel of Fig.~\ref{fig:Corr_MN1_theta} when $\theta_R$ is sizable while fixing $\theta_L = 10^{-15}$, thereby isolating the effect of the right-handed mixing. In both cases, the relic abundance initially increases as dark matter particles are gradually produced from the thermal bath through freeze-in processes. As the temperature decreases and the abundance grows, inverse reactions become increasingly relevant and partially counteract the production rate. This leads to a mild decrease in the abundance after it reaches its peak before eventually approaching a constant value at late times. 

In conclusion, for the model considered here, the dominant contributions arise from decay channels such as $h, H \rightarrow \overline{N}_1 \nu$ and $h, H \rightarrow N_1 \overline{\nu}$, as well as processes involving electroweak gauge bosons, including $W^+ \to N_1 e^+_i$, $W^- \to \overline{N}_1 e^-_i$, and $Z \rightarrow N_1 \overline{N}_1$. The relative importance of these channels is controlled by the mixing angles $\theta_L$ and $\theta_R$, which determine the strength of the interaction between the dark sector and the SM bath.

\subsection{Contribution to $N_{\rm eff}$ from $\nu_R$ production}
\label{sec:cosmo}
Since we are working in a Dirac neutrino framework, the light right-handed neutrinos $\nu_R$ contribute as additional relativistic species, thereby modifying the effective number of relativistic degrees of freedom in the early Universe, denoted as $\Neff$. This quantity is highly constrained from current cosmic microwave background observations~\cite{DESI:2024mwx}. Here we compute the contribution to the effective number of relativistic neutrino species $\Delta\Neff$, coming from $\nu_R$ production, which do not thermalise in the early Universe and are instead generated via freeze-in. Their production channels and corresponding squared decay amplitudes $|\mathcal{M}|^2$ are:
\begin{align}
\label{eq:vRprodamp}
    h \to \overline{N}_{L_i} \nu_{R_i} &: \;\;\; |\mathcal{M}|^2 = 2 g_{h \bar{N_i} \nu_i}^2 [m_h^2 - (m_{N_i} + m_{\nu_i})^2] \;, \\ \nonumber
    H \to \overline{N}_{L_i} \nu_{R_i} &: \;\;\; |\mathcal{M}|^2 = 2 g_{H \bar{N_i} \nu_i}^2 [m_H^2 - (m_{N_i} + m_{{\nu_i}})^2] \;.
\end{align}
where $g_{H \bar{N_i} \nu_i}$ and $g_{h \bar{N_i} \nu_i}$ are given in eq.~\eqref{eq:vertex}. To compute $\Delta N_{\rm eff}$ in freeze-in regime, we follow the method presented in Ref.~\cite{Luo_2021}. The Boltzmann equations for the new $\nu_R$-SM interactions are given as \citep{Luo_2020, Luo_2021},
\begin{align}
    \dot\rho_{\nu_R} + 4\left(\frac{\dot{a}}{a}\right)\rho_{\nu_R} & = C_{\nu_R} \label{eq:nuR_evol}\,, \\
    \dot\rho_{\rm SM} + 3\left(\frac{\dot{a}}{a}\right)(\rho_{\rm SM} + p_{\rm SM}) & = -C_{\nu_R}  \,,
\label{eq:SM_evol}
\end{align}
where $a$ is the cosmological scale factor, $\rho_{\rm SM}$ and $\rho_{\nu_R}$ are the energy densities of SM particles and $\nu_R$, respectively, $p_{\rm SM}$ is the pressure term of SM particles and $C_{\nu_R}$ denotes the collision term for the $\nu_R$-SM interaction. The collision term for the $1\xrightarrow{}2$ process that we consider for $\nu_R$ production can be written as,
\begin{align}
        C_{\nu_R} &= N_{\nu_R}\int E_{\nu_R} \left(\prod_i \frac{1}{(2\pi)^3}\frac{d^3 p_i}{2E_i}\right)(2\pi)^4\;\delta^4(p_1 - p_2 - p_3) |\mathcal{M}|^2 \left(\frac{1}{\exp(E_1/T_1 ) - 1}\right).
\end{align}
Here, $N_{\nu_R}=6$ is the number of $\nu_R (\overline{\nu}_R)$ flavours; $E_{\nu_R}$ is the energy of $\nu_R$; $p_i, E_i$ and $T_i$ are the momentum, energy and temperature of the $i$-th particle. $|\mathcal{M}|^2$ denotes the squared decay amplitude of the $1\xrightarrow{}2$ processes whose expressions are given in eq.~\eqref{eq:vRprodamp}.

We solve the Boltzmann eqs. \eqref{eq:nuR_evol} and \eqref{eq:SM_evol} numerically using a Monte-Carlo integration code\footnote{\href{https://github.com/xuhengluo/Thermal_Boltzmann_Solver}{https://github.com/xuhengluo/Thermal{\_}Boltzmann{\_}Solver}} \cite{Luo_2021}. The contribution of $\nu_R$ to $\Neff$ is by definition,
\begin{equation}
    \Delta \Neff = \frac{8}{7}\left(\frac{11}{4}\right)^{4/3} \frac{\rho_{\nu_R,0}}{\rho_{\gamma,0}},
\end{equation}
where $\rho_{\nu_R, 0}$ and $\rho_{\gamma,0}$ are the energy densities of $\nu_R$ and photons at some time after $e^\pm$-annihilation. We can rewrite this in terms of the corresponding temperature ratios at $T_{SM}=10~{\rm MeV}$ as \cite{Luo_2020, Luo_2021},
\begin{equation}
    \Delta \Neff = N_{\nu_R} \left( \frac{T_{\nu_R, 10}}{T_{\gamma, 10}} \right)^4.
\end{equation}
In our case, we obtain no significant contributions to $\Delta N_{\rm eff}$ ($\sim 10^{-17}$) for typical values of the masses of $N_1$ and $H$ and couplings of the interactions, considered in the parameter space of Fig. \ref{fig:Corr_MN1_theta}. This is consistent with the results shown in Ref. \cite{Luo_2021} for a similar case. 

\section{Summary and outlook}
\label{sec:concl}
In this work, we have presented a minimal model where Dirac neutrino masses are ``seeded'' by DM. In particular, we show that the lightest sterile neutrino that mediates neutrino mass generation at tree-level via the Dirac seesaw mechanism can simultaneously serve as viable DM candidate. The Dirac nature of neutrinos is enforced by a $\mathcal{Z}_6$ symmetry that is spontaneously broken to a residual $\mathcal{Z}_3$ symmetry once a real singlet scalar $\sigma$ acquires a vacuum expectation value. This residual symmetry forbids all Majorana neutrino mass operators, thereby ensuring that neutrinos remain Dirac particles.

We considered the minimal particle content capable of producing a phenomenologically viable dark matter scenario, consisting of three generations of $N_{L,R}$ and $\nu_R$. A setup with only two generations, which would predict one massless active neutrino, is excluded by X-ray constraints arising from the radiative decay channel $N \to \nu \gamma$. Focusing on the lightest sterile neutrino $N_1$ as the dark matter candidate, we studied its non-thermal production via freeze-in from the decays of SM particles as well as the additional scalar $H$. At the same time, we ensured that the lifetime of $N_1$ exceeds the age of the Universe, rendering it effectively stable on cosmological timescales. Our parameter scan reveals two qualitatively distinct regimes. When production processes controlled by the left-handed mixing angle $\theta_L$ dominate, the allowed parameter space is strongly constrained by X-ray observations, leading to viable dark matter masses in the lower mass range ($M_{N_1} \sim 10^{-4}\,\mathrm{GeV}$). In contrast, when production is governed primarily by the right-handed mixing angle $\theta_R$, the X-ray limits are absent, since $\theta_R$ does not contribute to the $N\to\nu\gamma$ decay. This allows for viable dark matter at higher sterile neutrino masses ($M_{N_1} > 10^{-2}\,\mathrm{GeV}$). We also examined the cosmological implications of the model, in particular the contribution to the effective number of relativistic species, $\Delta N_{\rm eff}$. We find that the freeze-in production of right-handed neutrinos $\nu_R$ yields a negligible contribution across the entire parameter space, ensuring consistency with current cosmic microwave background constraints~\cite{DESI:2024mwx}.

Overall, our setup provides the simplest framework, in terms of field content and symmetries, in which a \emph{dark}-sector induces non-zero Dirac neutrino masses. Several alternative approaches have been explored in the literature, however they have several distinctions in the particle and symmetry content, as well as in the nature of the DM candidate considered. For instance, Ref.~\cite{CentellesChulia:2016rms} employed a $\mathcal{Z}_4 \otimes \mathcal{Z}_2$ symmetry and Ref.~\cite{Borboruah:2024lli} employed a $\mathcal{Z}_4 \otimes \mathrm{U(1)_{L}}$ symmetry to forbid Majorana neutrino masses, the bare Dirac Yukawa term ($\overline{\ell_L} \tilde{\Phi} \nu_R$), and to stabilize the DM candidate. Both works extended the particle content with additional scalar fields whose lightest component played the role of weakly interacting massive particle (WIMP) DM. Ref.~\cite{Reyimuaji:2024kqs} employed a $\mathcal{Z}_3 \otimes \mathcal{Z}_2$ symmetry considered sterile neutrinos transforming as $\mathrm{SU(2)_L}$ doublets, which behave as WIMP DM produced through freeze-out. Compared to our scenario where the DM lies in the sub-GeV range, the DM candidates in these WIMP scenarios lies at the TeV. 

\begin{acknowledgments}
We thank Rahul Srivastava for valuable discussions and Dhanashree Pathe for help with the numerical code. J.A. is supported by the National Science Centre, Poland (research grant No. 2021/42/E/ST2/00031). The work of K.D. is partially supported by the grant BPI/STE/2021/1/00033/U/00001. A.B. is supported by the PhD grant UI/BD/154391/2023 from the Fundação para a Ciência e a Tecnologia (FCT, Portugal). The work of A.B. and F.R.J. is also supported  by the \href{https://doi.org/10.54499/UID/00777/2025}{FCT project UID/00777/2025}.

\end{acknowledgments}

\appendix

\section{Scalar sector}
\label{sec:scalar}
As shown in Table~\ref{tab:model}, the scalar sector of our model contains the SM Higgs doublet $\Phi$ and a real singlet $\sigma$, which we define as,
\begin{align}
\Phi&=\begin{pmatrix}
\phi^{+} \\
\phi^0
\end{pmatrix}= \frac{1}{\sqrt{2}}  \begin{pmatrix}
 \sqrt{2} \phi^{+} \\
 v_\phi + \phi_{\text{R}} + i \phi_{\text{I}}
\end{pmatrix} \; ; \; 
\sigma = \frac{v_\sigma + \sigma_{\text{R}}}{\sqrt{2}} \; .
\label{eq:scalars}
\end{align}
The most general scalar potential is given by,
\begin{align}
    V &=  m_{\Phi}^2 \left(\Phi^\dagger \Phi\right) + m_\sigma^2 \sigma^2 + \frac{\lambda_{\Phi}}{2} \left(\Phi^\dagger \Phi\right)^2 + \frac{\lambda_{\sigma}}{2} \sigma^4 + \lambda_{\Phi \sigma} \left(\Phi^\dagger \Phi\right) \sigma^2 \; .
\end{align}
To ensure that it is bounded from below, the quartic couplings must fulfill the conditions~\cite{Ivanov:2018jmz},
\begin{equation}
    \lambda_\Phi \ge 0 ,\; \lambda_\sigma \ge 0 ,\; \text{and} \;
    \lambda_{\Phi \sigma} \ge -\sqrt{\lambda_\Phi \lambda_\sigma} \;.
\end{equation}
There are two neutral CP-even Higgs scalars $h$ and $H$ (from the mixing between $\phi_R$ and $\sigma_R$) with masses
\begin{align}
m_{h,H}^2 &= \frac{1}{2} \left( v_\phi^2 \lambda_\Phi + v_\sigma^2 \lambda_\sigma \pm \sqrt{ \left( v_\sigma^2 \lambda_\sigma - v_\phi^2 \lambda_\Phi \right)^2 + 4 v_\phi^2 v_\sigma^2 \lambda_{\Phi \sigma}^2} \right) \; . 
\end{align}
The $h$ and $H$ mass eigenstates are related to $\phi_R$ and $\sigma_R$ through the $2 \times 2$ orthogonal matrix:
\begin{equation}
\begin{pmatrix}
h \\
H
\end{pmatrix}
 = \begin{pmatrix}
\cos{\alpha} & -\sin{\alpha} \\
\sin{\alpha} & \cos{\alpha}
\end{pmatrix} 
\begin{pmatrix}
\phi_R \\
\sigma_R
\end{pmatrix} \; ,
\label{eq:scalarmixing}
\end{equation}
where $\alpha$ is the mixing angle. We can express the potential terms $\lambda_{\Phi}$, $\lambda_{\sigma}$, and $\lambda_{\Phi\sigma}$ in terms of $\alpha$, $m_{h}$ and $m_{H}$ as follows:
\begin{align}
\lambda_\Phi &=\frac{c_\alpha^2 m_h^2 + s_\alpha^2 m_H^2}{v_\phi^2} \; , \nonumber \\ 
\lambda_\sigma &=\frac{s_\alpha^2 m_h^2 + c_\alpha^2 m_H^2}{v_\sigma^2} \; , \nonumber \\
\lambda_{\Phi\sigma} &=\frac{c_\alpha s_\alpha (m_H^2-m_h^2)}{v_\phi v_\sigma}\; ,
\label{eq:lambda}
\end{align}
where $c_\alpha=\cos{\alpha}$ and $s_\alpha=\sin{\alpha}$. 

\section{Relevant interaction vertices}
\label{sec:vertex}
To find the relic density of $N_1$, the cross-sections of all the dominant production processes must be computed. Their analytical expressions have been presented in Ref.~\cite{Biswas:2016bfo} for a U$(1)_{B-L}$ model and can be used separately for $N_1$ and $\overline{N}_1$ here. The relevant interaction vertices for the processes given in eq.~\eqref{eq:process} are:
\begin{align}
\label{eq:vertex}
    g_{h \overline{N}_1 \nu_1}&=\frac{1}{\sqrt{2}} (- Y_{\nu}^{11} c_{\alpha} \sin{\theta_{L1}} \sin{\theta_{R1}}  - Y_{\sigma}^{11} s_{\alpha} \cos{\theta_{L1}} \cos{\theta_{R1}}) \; , \\\nonumber
    g_{H \overline{N}_1 \nu_1}&=\frac{1}{\sqrt{2}}(- Y_{\nu}^{11} s_{\alpha} \sin{\theta_{L1}} \sin{\theta_{R1}} + Y_{\sigma}^{11} c_{\alpha} \cos{\theta_{L1}} \cos{\theta_{R1}} ) \; , \\\nonumber
    g_{h N_1 \overline{\nu}_1}&=\frac{1}{\sqrt{2}} (Y_{\nu}^{11} c_{\alpha} \cos{\theta_{L1}} \cos{\theta_{R1}}  + Y_{\sigma}^{11} s_{\alpha} \sin{\theta_{L1}} \sin{\theta_{R1}} ) \; , \\\nonumber
    g_{H N_1 \overline{\nu}_1}&=\frac{1}{\sqrt{2}}(Y_{\nu}^{11} s_{\alpha} \cos{\theta_{L1}} \cos{\theta_{R1}} - Y_{\sigma}^{11} c_{\alpha} \sin{\theta_{L1}} \sin{\theta_{R1}} ) \; .
\end{align}
All terms proportional to the sine of the neutrino mixing angles are negligible compared to those proportional to the cosine. 

\end{document}